\documentclass[letterpaper]{sigchi}
\usepackage[pass]{geometry}
\usepackage{color}
\usepackage{graphicx}
\usepackage{url}
\usepackage{adjustbox}
\usepackage{booktabs}
\usepackage{array}
\usepackage{balance} 
\usepackage{paralist} 
\usepackage[caption=false]{subfig}
\newcolumntype{M}{>{\centering\arraybackslash}m{0.6in}}
\usepackage{multirow}

\begin{document}

\title{Social Media Image Analysis for Public Health}


\numberofauthors{3}
\author{%
  \alignauthor{Kiran Garimella\\
    \affaddr{Aalto University}\\
    \affaddr{Helsinki, Finland}\\
    \email{kiran.garimella@aalto.fi}}\\
  \alignauthor{Abdulrahman Alfayad\\
    \affaddr{Carnegie Mellon University}\\
    \affaddr{Doha, Qatar}\\
    \email{afayad@qatar.cmu.edu}}\\
  \alignauthor{Ingmar Weber\\
    \affaddr{Qatar Computing Research Institute}\\
    \affaddr{Doha, Qatar}\\
    \email{iweber@qf.org.qa}}\\
}

\maketitle

\begin{abstract}
Several projects have shown the feasibility to use \emph{textual} social media data to track public health concerns, such as temporal influenza patterns or geographical obesity patterns. In this paper, we look at whether geo-tagged \emph{images} from Instagram also provide a viable data source. 
Especially for ``lifestyle'' diseases, such as obesity, drinking or smoking, images of social gatherings could provide information that is not necessarily shared in, say, tweets. In this study, we explore whether (i) tags provided by the users and (ii) annotations obtained via automatic image tagging are indeed valuable for studying public health. 
We find that both user-provided and machine-generated tags provide information that can be used to infer a county's health statistics. Whereas for most statistics user-provided tags are better features, for predicting excessive drinking machine-generated tags such as ``liquid'' and ``glass'' yield better models. This hints at the potential of using machine-generated tags to study substance abuse. 
%
\end{abstract}



\section{Introduction}\label{sec:introduction}

The annual cost of ``lifestyle diseases'' in the U.S.\ is estimated at up to a \emph{trillion} U.S.\ dollars.
In order to target these diseases and induce required lifestyle changes it is useful to first gain a better understanding of the phenomenon through (i) identifying the population groups most affected, (ii) understanding the interaction between environment, population and other variables, and (iii) tracking changes over time to monitor the effect of any campaign or interventions.

Over the last couple of years, social media has emerged as a viable data source to use for large-scale public health studies. Using social media seems particularly appropriate for studying lifestyle diseases, such as obesity, as more and more of people's life and social interaction is publicly shared online. 

One type of social media data that, to the best of our knowledge, has not been tapped into for large-scale public health studies is \emph{images}. More than 1.8 billion images are shared online every day
\footnote{\url{http://read.bi/1ouGzgQ}, last accessed on July 13, 2015.},
 (compared to 500M tweets per day), making rich media content a prolific data source.

In addition to its sheer size, social media image data might contain complementary data that is not shared in textual status updates. For example, after or even during a social gathering it is often customary to share images taken with a smart phone, rather than post a textual summary of the event. 
Images shared might or might not be annotated by the user. Until recently, automatically processing the \emph{content} of images was beyond the realm of technical feasibility. Due to advances in deep learning~\cite{karpathy2014deep}, it has become possible to automatically understand objects in images and generate textual descriptions. 
%
In this paper, we look at whether the $content$ of images shared on Instagram provides health-related information, and, in particular, whether machine-generated tags add value over user-provided ones.

The use of automatic image annotation tools holds a range of potential advantages one of which is language agnosticism. Here, ``language'' refers both to English vs.\,~say, Spanish but it also refers to different and inconsistent conventions of what users might label as, say, ``\#healthy''. 
Another advantage is the inference of ``hidden'' labels that an image owner would not assign. Studies have shown that a large fraction of young social media users have posted a picture depicting substance use~\cite{morgan2010image}. Such images about alcohol or drug intake may not be explicitly labeled as, say, ``\#alcohol'' and hence difficult to track using text-based methods.
The insights obtained from this study could be of use in designing systems for understanding and monitoring substance abuse and other stigmatized behaviors.

We find that both user-provided and machine-generated labels for geo-tagged Instagram images hold valuable information in predicting county-level health statistics. Though for most variables user-provided labels outperform machine-generated ones, the opposite holds for ``Excessive Drinking''.
%



\section{Related Work}\label{sec:related}

\textbf{Social media data for public health analysis}
%
Traditional methods to track public health metrics are based on pooling data from doctors, pharmacies and other sources which comes with certain drawbacks related to latency and cost. For tracking lifestyle diseases, where latency is not an issue, survey-based approaches have difficulties in collecting a ``holistic'' view of a patient that includes many facets of their lifestyle.

Recent studies have shown that a large scale, real time, non-intrusive monitoring can be done using social media to get aggregate statistics about the health and well being of a population~\cite{dredze12is,sarker15jbi,kostkova13www}. Twitter in particular has been widely used in studies on public health~\cite{pauldredze11icwsm,prier2011identifying,parkeretal13asonam,Kostkova15TwitterSocioscope}, due to its vast amount of data and the ease of availability of data. 

Culotta~\cite{culotta14chi} and Abbar et al.~\cite{abbaretal15chi}
	used Twitter in conjunction with psychometric lexicons such as LIWC and PERMA to predict county-level health statistics such as obesity, teen pregnancy and diabetes. 
Paul et al.~\cite{pauldredze14pone} make use of Twitter data to identify health related topics and use these to characterize the discussion of health online. 
Mejova et al.~\cite{mejovaetal15dh} use Foursquare and Instagram images to study food consumption patterns in the US, and find a correlation between obesity and fast food restaurants.





Social media has also enabled large-scale studies linking lifestyle and  health data at an $individual$ level. For example, Sadilek et al.~\cite{sadilekkautz13wsdm}, build a classifier to identify the health of a user based on their Twitter usage. 
Many more ``hidden'', yet important conditions such as depression~\cite{de2013predicting,balani2015detecting,andalibi2015depression}, sleep problems~\cite{mciveretal15jmir}, eating disorders~\cite{walker2015facebook}, and substance use~\cite{salimianetal14pa} have been studied using social media data. 
%
%
%
Our study is different from the ones discussed above in that we propose the use of image data to study public health. 

Abdullah, et al.~\cite{abdullah2015collective} use smile recognition from images posted on social media to study and quantify the overall societal happiness. Andalibi et al.~\cite{andalibi2015depression} study depression related images on Instagram and ``establish[ed] the importance of visual imagery as a vehicle for expressing aspects of depression". In our work, we study if the use of image recognition techniques helps in understanding a broader range of health related issues.






\textbf{Image recognition, Deep learning}
Almost all the methods above rely on textual content though images and other rich multimedia form a major chunk of content being generated and shared in social media. 
Automatic image annotation has greatly improved over the last couple of years, owing to the development in deep learning~\cite{hinton2006fast}. 
Object recognition~\cite{wu2015deep} and image tagging~\cite{karpathy2014deep} have become possible because of these new developments, e.g.\ Karpathy et al.~\cite{karpathy2014deep} use deep learning to produce descriptions of images, which compete with (and sometimes beat) human generated labels.  
A few studies already make use of these advances to identify~\cite{kawanoyanai15mta} and study~\cite{sudoetal14ubicomp} food consumption from pictures. 
%














\section{Data Collection}\label{sec:data}

The social media data used in our analysis consists of geo-referenced Instagram images (mostly taken at restaurants, related to food) obtained by Mejova et al.~\cite{mejovaetal15dh}. From this data, we are using the November subset consisting of 18M images. The choice of a smaller subset was randomly made to cater to the hard rate-limit restrictions imposed by Imagga and the choice for food related images was to maintain topical focus.

Each image is associated with various types of meta data, such as geographic location (latitude, longitude), user-provided hashtags, time stamp and comments. Using the (latitude, longitude), we mapped an image to a U.S.\ county (represented by a unique FIPS code) using an API provided by Federal Communication Commission.
We then considered the top 100 counties in terms of image count and sampled 2,000 images uniformly at random from each of these 100 counties, due to computational restrictions imposed by the image annotation API. 




We make use of two types of meta data obtained from images in this study: (i) machine-generated tags from a commercial service, and (ii) user-provided textual tags (hashtags).

\subsection{Machine-Generated Tags}
\begin{figure}
   \centering
   \includegraphics[width=0.5\textwidth,height=0.3\textwidth]{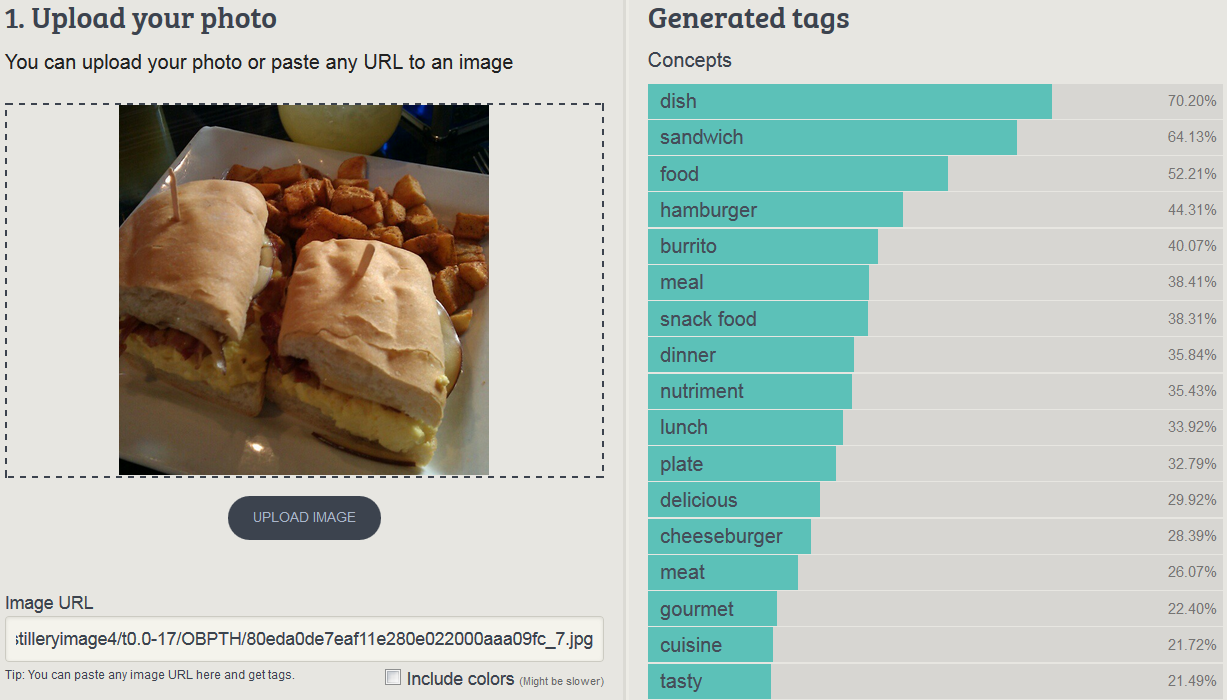}
      \caption{Example output for the Imagga Auto-Tagging Demo for one of the Instagram images in our data set. Each tag is associated with a confidence score (shown on the right).}
    \label{fig:imagga-example}
    \vspace{-\baselineskip}
\end{figure}

Each image was tagged using a state-of-the-art commercial image recognition tool from Imagga.com.\footnote{\url{http://docs.imagga.com/#auto-tagging}} 
Figure~\ref{fig:imagga-example} shows an example image along with the tags and their confidence. 
For 189,293 of our 200,000 images we managed to obtain at least one tag via Imagga. For the remaining 10,707 images the Instagram image was no longer publicly available, either due to deletion, changes in privacy settings, or account suspension by Instagram.\footnote{The Imagga tags dataset is available for download at \url{https://users.ics.aalto.fi/kiran/imagga_tags_data}} 

Imagga API returns, for each image a set of tags along with a `confidence' score. We only considered those tags with a confidence score above 20\%. 
We experimented with different threshold values in the range of [5,60]. Based on the macro-averaged correlation for the nine different indices, we selected a global value of 20\%. Though different threshold values might lead to slightly better results for a given index, we preferred not to explore this route to avoid the risk of overfitting.
Tags appearing in less than 10 distinct counties were also ignored, leaving us with 1,750 distinct tags. Using this tag set, we created a representation of the tags used in each of the 100 counties by counting how many images were assigned a particular tag by Imagga.
%
%
%
Imagga is not a restaurant- or food-specific annotation tool and its model is trained on general images. 
The ten most commonly returned machine annotations for the images in our data set were `food', `delicious',`cocktail',`people',`gold'
, `healthy', `adult', `caucasian', `person', `kitchen'. We observe that the tags  contain references to food, restaurants and people. In general, the Imagga generated tags include a wide range of information from identifying types of food (`spaghetti'), drinks (`cappuccino'), objects (`stringed instrument'), people (`male'), and actions (`drunkard').





\subsection{User-Provided Tags}

Many images are provided with textual tags by the user uploading the image. For example, the image shown in 
Figure~\ref{fig:imagga-example}
comes with the tags \#breakfastsandwich, \#brunch, and \#amazingpotatoes. Note that images might or might not have textual tags assigned by the user (in our dataset, 49\% of the images have at least one tag). To have an equal size comparison to the machine-annotated data set, we sampled 200k images with at least one user-provided tag, 2,000 images for each of the 100 counties.
Only user-provided tags appearing in at least 10 distinct counties were kept, leaving us with 5,865 distinct tags.




\subsection{Demographic variables}
As a baseline, we also obtained a list of demographic variables from the County Health Rankings website.\footnote{\url{http://www.countyhealthrankings.org/rankings/data}} For each of the 100 counties we used same the demographic features that were used in Culotta~\cite{culotta14chi}, representing information related to (i) age distribution, (ii) racial distribution, and (iii) income. 
%
%
%
%
The ground-truth offline data on public health was obtained from the same County Health Rankings site.
This data contains a wealth of health-related variables ranging from ``Teen births'' to ``Diabetic monitoring''. 
Given that our image data was obtained from \emph{food} locations, we decided to focus our analysis on the following nine variables: 
(i) $\%\ Smokers$, (ii) $\%\ Obese$, (iii) $Food\ environment\ Index$, (iv) $\%\ Physically\ inactive$, (v) $\%\ Excessive\ drinking$, (vi) $\% Alcohol-impaired\ driving\ deaths$, (vii) $\%\ Diabetics$, (viii) $\% Food\ insecure$, and (ix) $\%\ Limited\ access\ to\ healthy\ food$.
%

\section{Methods}\label{sec:methods}

To test whether social media image data helps in studying public health, we perform regression to predict each of the nine health-related variables (dependent variables) using the image meta data features described above (independent variables). 
Given the high-dimensionality of our feature vectors relative to the number of validation points (N=100 counties), we use ridge regression to avoid over fitting (with a smoothing parameter of $\alpha$=0.1 as in \cite{culotta14chi}).
We perform 10 fold cross-validation, training a model on 90 counties and testing it on the remaining 10 to evaluate the accuracy of the regression.

For each county, the number of images containing a tag was used to create a county x tags matrix (size 100 x 1,750 for Imagga, 100 x 5,865 for user tags). 
We normalized each row (county) in this matrix using Euclidean norm in order to remove differences in image counts across counties. 
We also tried out various extensions of the text tags using external dictionaries such as LIWC~\cite{pennebaker07liwc} and PERMA~\cite{seligman12perma}. We do not report results for these here as they did not lead to improvements, neither applied to human-generated nor to machine-generated tags.

To measure the accuracy, we used (i) Pearson's r - Correlation between the actual and predicted values (larger values are better), and (ii) Symmetric Mean Absolute Percentage Error (SMAPE, smaller values are better).

\section{Results}\label{sec:results}

Table~\ref{tab:results} summarizes the prediction performance for different combinations of (i) user-provided tags as features (``U''), (ii) machine-generated tags as features (``I''), and (iii) county-level demographics (``D''). We observe that using user-provided tags in conjunction with demographic features helps significantly improve the performance over just using demographic features, both in terms of Pearson's r and SMAPE.

\begin{table*}
\centering
\caption{Prediction performance for nine health statistics across 100 counties in a 10-fold cross validation setting. Feature sets used were (U)ser tags, (I)magga tags, (D)emographics and combinations of these. Statistical significant improvement over the demographics-only baseline was tested using a Fisher r-z transform paired test between the two correlations $^{*}=.05$, $^{**}=.01$, $^{***}=.001$.\label{tab:results}}
{\resizebox{\linewidth}{!}{
\begin{tabular}{|c|c|c|c|c|c|c||c|c|c|c|c|c|c}
\hline
& \multicolumn{6}{|c|} {Pearson's r} & \multicolumn{6}{||c|} {SMAPE} \\
\hline
 & U & I & D & U+D & I+D & U+I+D & U & I & D & U+D & I+D & U+I+D \\
\hline
Smokers & 0.55 & 0.53 & 0.72 & 0.73 & \textbf{0.75} & 0.73 & 8.1 & 8.7 & 7.1 & 6.8 & \textbf{6.7} & 6.9 \\
Obese & 0.42 & 0.48 & 0.81 & \textbf{0.84$^*$} & 0.82 & 0.84 & 6.5 & 6.2 & 4 & \textbf{3.7} & 3.9 & 3.7 \\
Food Env. Index & 0.54 & 0.45 & 0.87 & \textbf{0.91$^{***}$} & 0.87 & 0.90 & 4.2 & 4.3 & 2.4 & \textbf{2} & 2.4 & 2 \\
Physically active & 0.46 & 0.58 & 0.70 & \textbf{0.79$^{*}$} & 0.74 & 0.79 & 7.7 & 7.4 & 5.9 & \textbf{4.8} & 5.6 & 4.9 \\
Excessive Drinking & 0.22 & \textbf{0.49$^{**}$} & 0.38 & 0.33 & 0.48 & 0.38 & 6.2 & \textbf{5.2} & 6.1 & 6.3 & 5.7 & 6 \\
Alcohol Impaired & \textbf{0.51$^{**}$} & 0.26 & 0.25 & 0.43 & 0.28 & 0.44 & \textbf{7.9} & 8.9 & 9.1 & 8.5 & 8.9 & 8.2 \\
Diabetic & 0.34 & 0.40 & 0.77 & \textbf{0.83$^{**}$} & 0.78 & 0.82 & 7.7 & 7.2 & 5.1 & \textbf{4.5} & 4.9 & 4.6 \\
Food insecure & 0.37 & 0.38 & 0.84 & 0.86 & 0.84 & \textbf{0.87} & 7.8 & 7.2 & 4.4 & \textbf{4.2} & 4.5 & 4.3 \\
Limited access & \textbf{0.61$^{*}$} & 0.40 & 0.46 & 0.58 & 0.48 & 0.58 & \textbf{19.7} & 22.6 & 21.5 & 20.5 & 20.9 & 20.8 \\
\hline
\end{tabular}}}
\vspace{-\baselineskip}
\end{table*}

To understand the performance of the models in more detail, we decided to zoom in on the four variables with the largest absolute improvement in Pearson's r over the demographics-only baseline: physically active (U+D), excessive drinking (I), alcohol impaired driving deaths (U), and limited access to healthy food (U).

\begin{figure}[h]
\begin{minipage}{.49\linewidth}
\centering
\subfloat[]{\label{}\includegraphics[width=\textwidth, height=\textwidth, clip=true, trim=10 90 30 90]{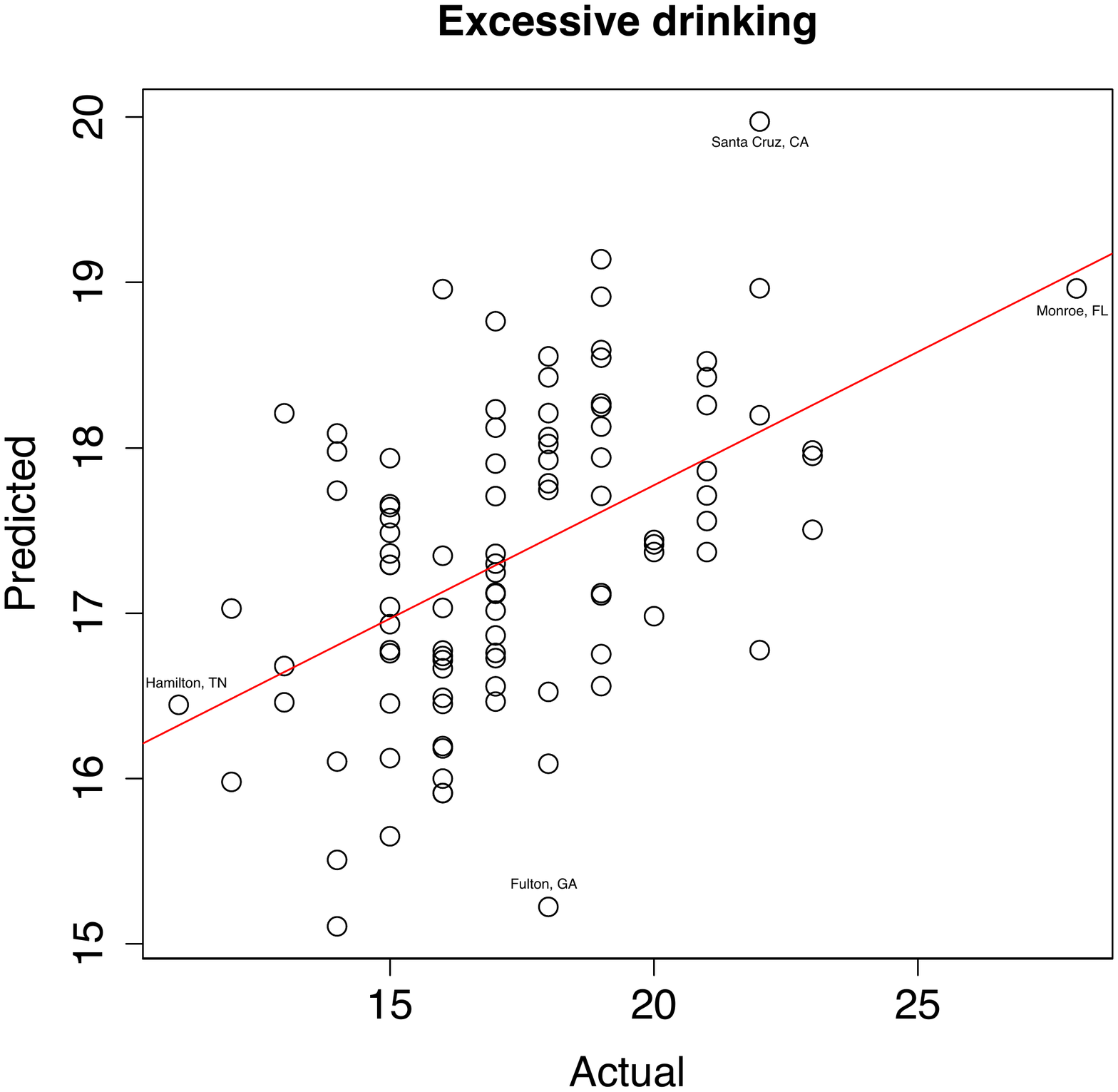}}
\end{minipage}%
\begin{minipage}{.49\linewidth}
\centering
\subfloat[]{\label{}\includegraphics[width=\textwidth, height=\textwidth, clip=true, trim=10 90 30 90]{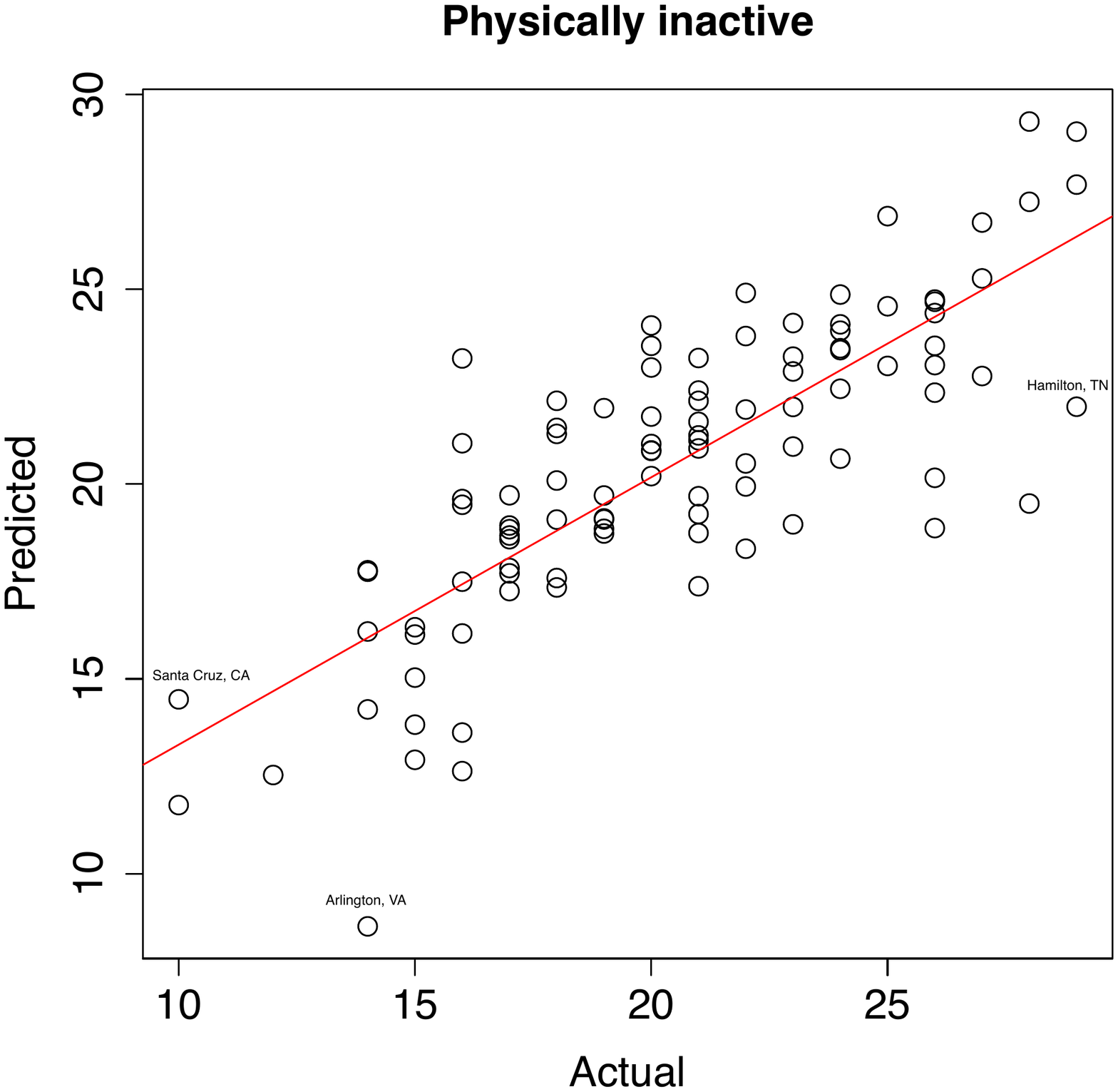}}
\end{minipage}\par\medskip
\begin{minipage}{.49\linewidth}
\centering
\subfloat[]{\label{}\includegraphics[width=\textwidth, height=\textwidth, clip=true, trim=10 90 30 90]{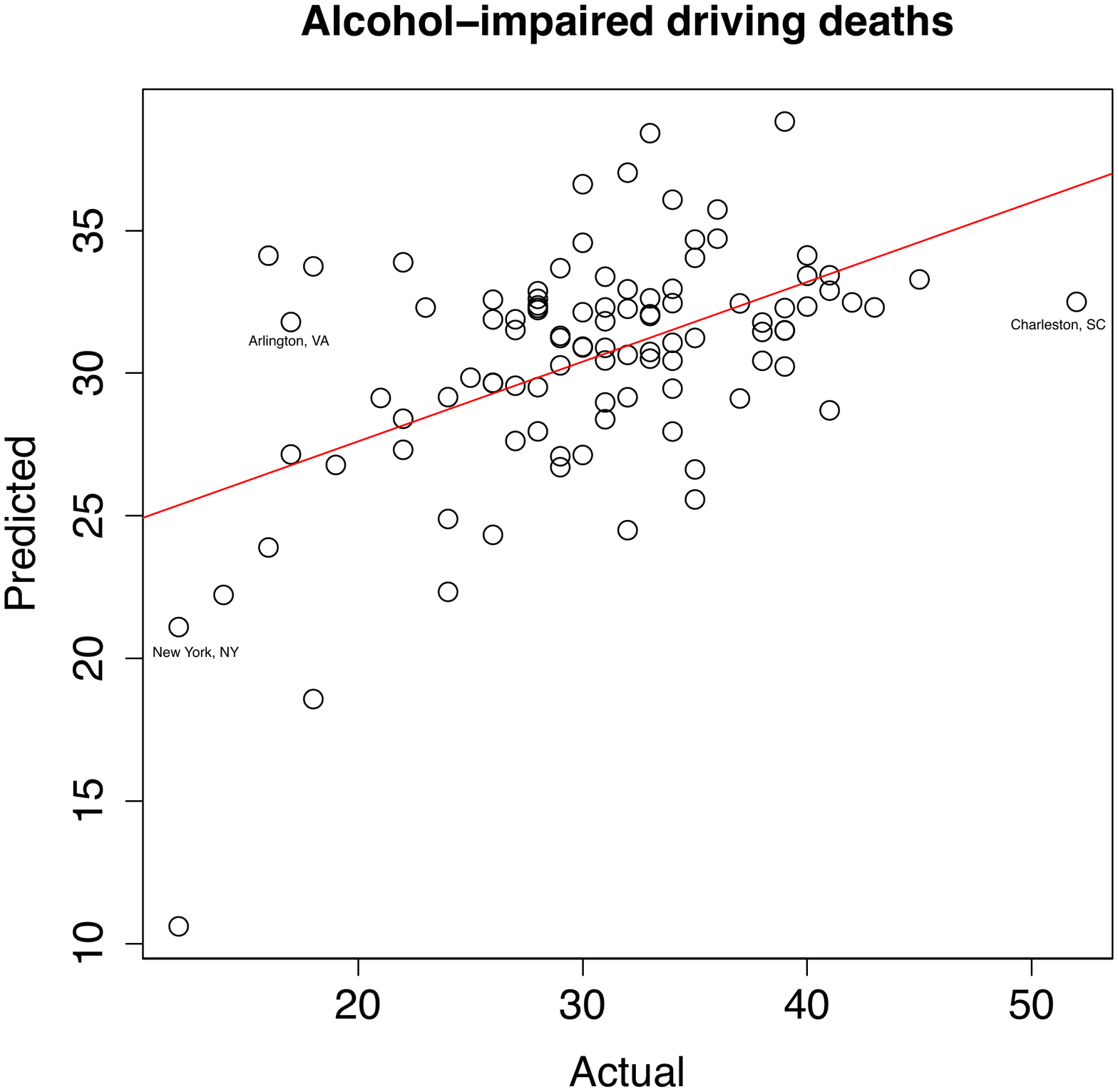}}
\end{minipage}%
\begin{minipage}{.49\linewidth}
\centering
\subfloat[]{\label{}\includegraphics[width=\textwidth, height=\textwidth, clip=true, trim=10 90 30 90]{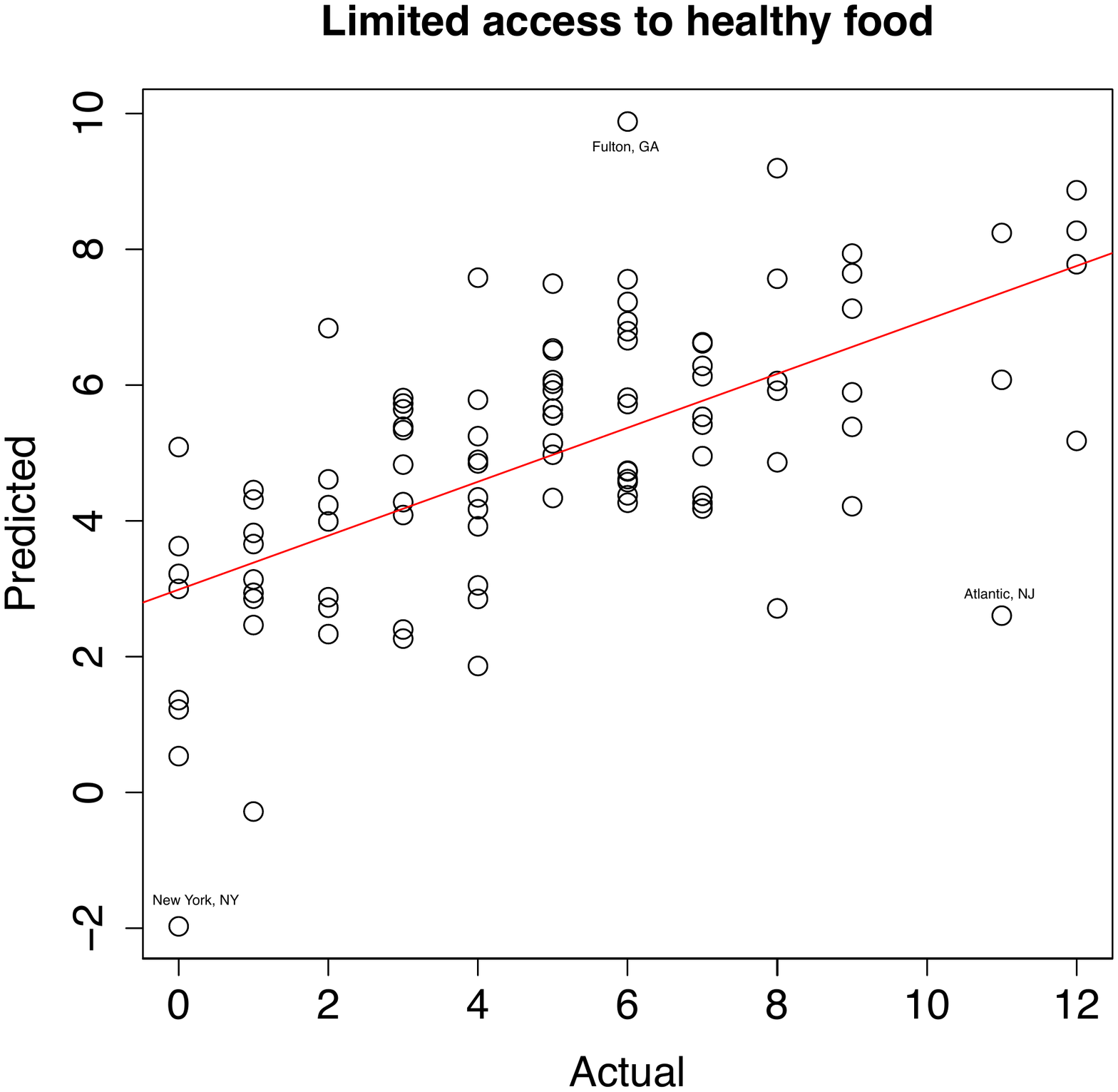}}
\end{minipage}\par\medskip

\caption{Predicted vs.\ actual health statistics value using (a) Imagga tags for predicting excessive drinking, (b) User-provided tags + demographics for predicting physically inactive, (c) User-provided tags for predicting alcohol-impaired driving deaths, and (d) User-provided tags for predicting limited access to healthy food.}
\label{fig:scatter}
\vspace{-\baselineskip}
\end{figure}

Figure~\ref{fig:scatter} shows scatter plots for these four health statistics. Each dot corresponds to a county with its actual health value on the x-axis and the predicted value, using the best-performing model, on the y-axis. As it might be of interest to domain experts, we labeled some of the outliers on the scatter plot with the name of the area.

According to our data, Arlington, VA ``looks'' healthier than it actually is, whereas the opposite holds for Hamilton, TN. In future work, it would be interesting to see what differences exist between such counties. 


\begin{figure}[h]
\begin{minipage}{.48\linewidth}
\centering
\subfloat[]{\label{}\includegraphics[width=\textwidth, height=\textwidth, clip=true, trim=20 150 30 180]{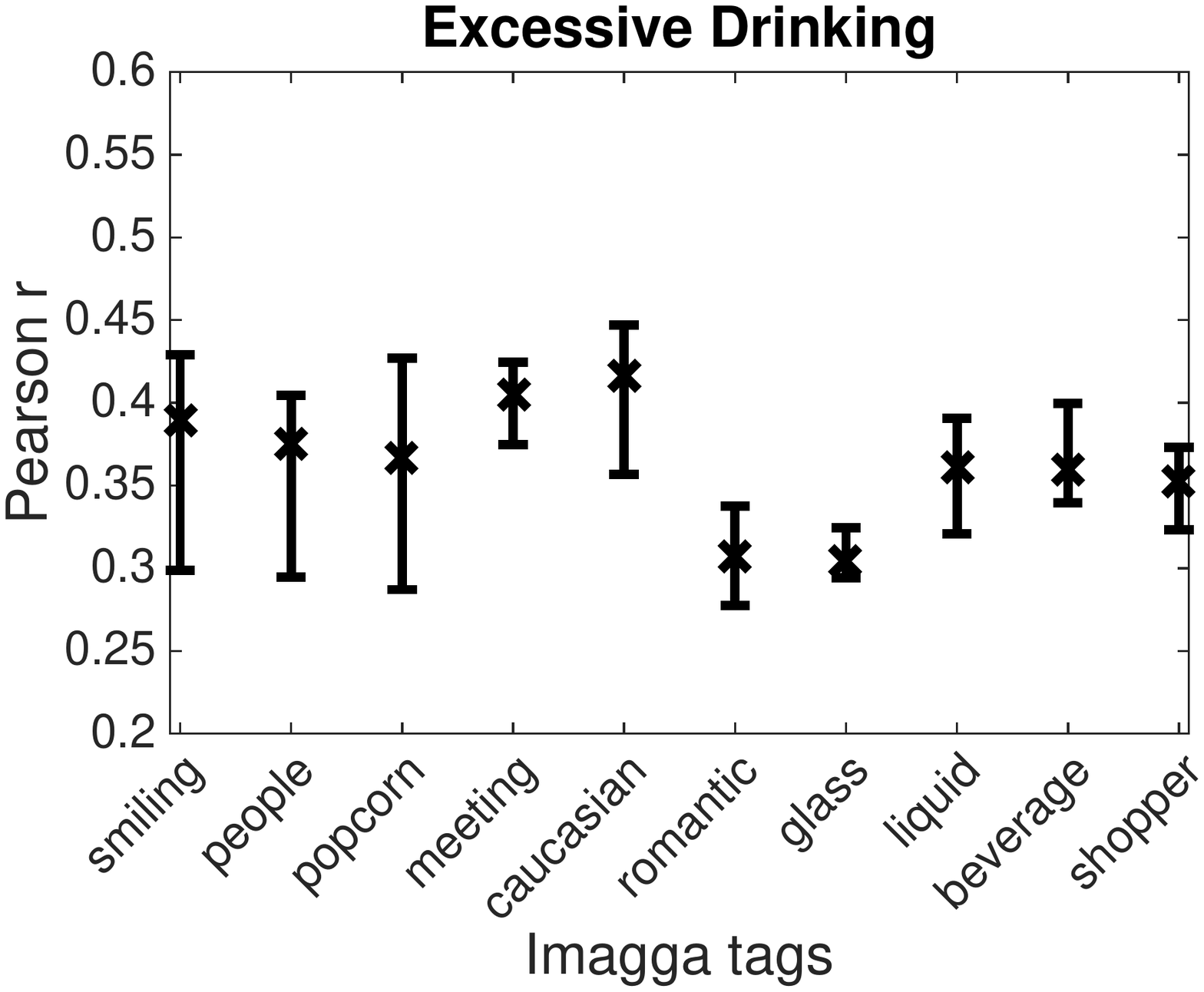}}
\end{minipage}%
\hspace{0.01\textwidth}
\begin{minipage}{.48\linewidth}
\centering
\subfloat[]{\label{}\includegraphics[width=\textwidth, height=\textwidth, clip=true, trim=20 150 30 180]{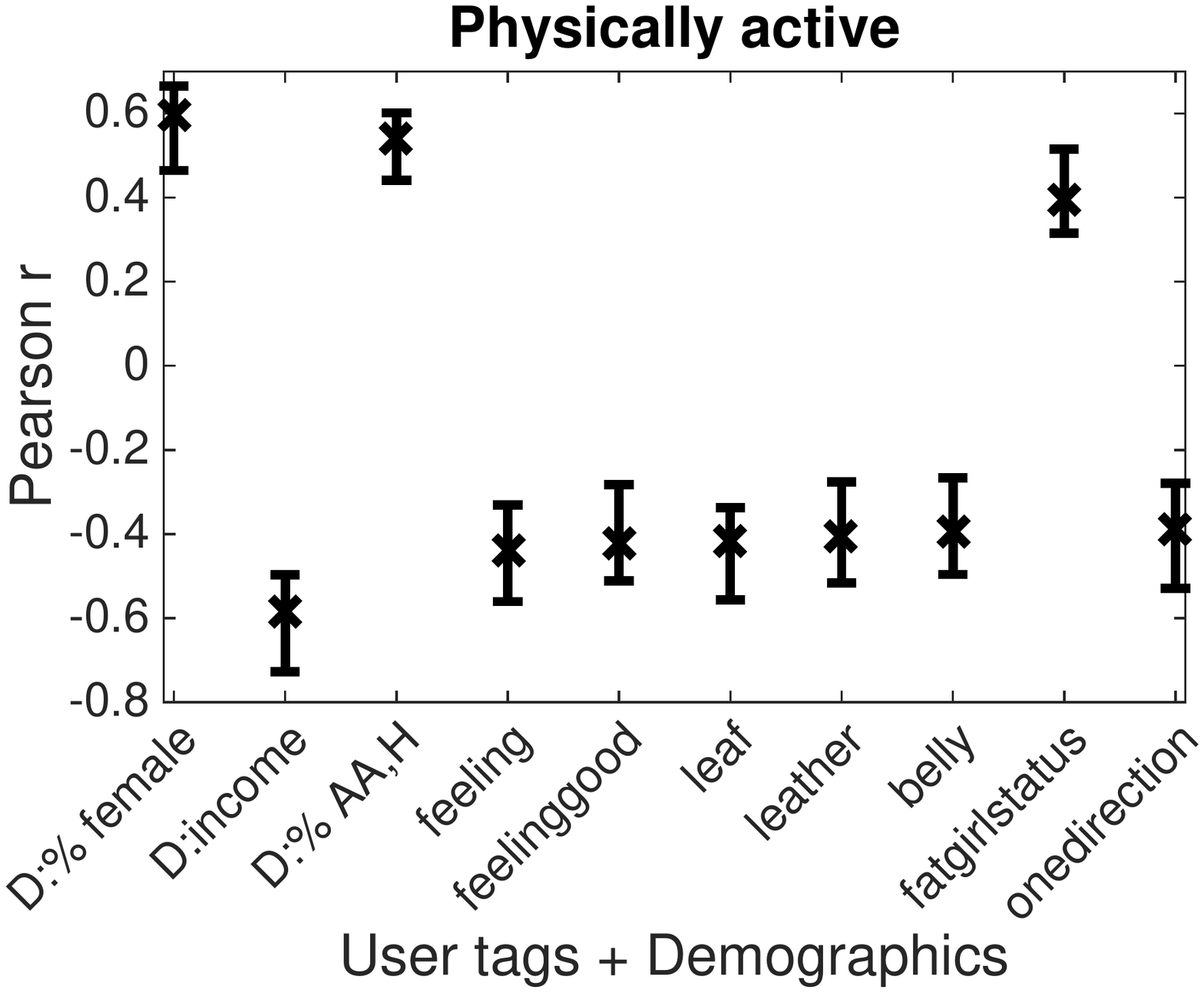}}
\end{minipage}\par\medskip

\caption{Feature analysis, showing the top ten features with the highest absolute value of correlation, for Imagga tags for predicting excessive drinking, (b) User-provided tags + demographics for predicting physically inactive.
In (b), features with ``D:'' indicate demographic features, ``D: AA,H'' indicates demographic features pertaining to African Americans and Hispanics. Error bars indicate 95\% confidence interval.}
\label{fig:feature_analysis}
\vspace{-\baselineskip}
\end{figure}

To understand \emph{why} a particular model works, we performed a feature analysis for the same set of four statistics. In each case we computed the top 10 features with the highest (absolute) value of correlation.
Figure~\ref{fig:feature_analysis} shows the results.

Among the machine-derived tags most predictive of excessive drinking (Fig.~\ref{fig:feature_analysis} (a)), ``liquid'',``beverage'' and ``meeting'' (c.f.\ ``party'') and others make intuitive sense.
These tags are especially interesting since such tags might not be part of human annotations. For the user-provided tags predictive of physical inactivity (Fig.~\ref{fig:feature_analysis} (b)), the hashtag \#fatgirlstatus makes sense. It is used in a self-ironic manner to refer to food obsession and related topics.
In addition to these intuitive features, there are also many that require further investigation. As the goal of this study was to test the general feasibility of using \emph{images} for public health studies, rather than to fully explain the social dynamics behind one particular health issue, we leave this for future work.

\section{Conclusions}
In this paper, we studied if social media image data can be used to understand public health. We used both user-provided as well as machine-generated tags to see if they can predict county-level health statistics. Our study relies on recent advances in deep learning for image analysis, a very active field that is undergoing constant advances.
We show that both user-provided and machine-generated tags provide information that improves over models using only demographic information. Whereas in most cases, models based on user-provided tags outperform machine-generated ones, for statistics on excessive drinking the opposite holds. Machine-generated tags such as ``glass'', ``liquid'' and ``beverage'' (Figure~\ref{fig:feature_analysis}) are all found to be good predictors, and such tags might not be part of human annotations. 
Machine-generated tags might prove of particular value in the case of stigmatized behaviors such as substance abuse, where explicit hints are unlikely to be added by the image owner. 
This could also be used in designing interfaces that could identify and help users in need of support.




\section{Acknowledgements.}
Kiran Garimella has been supported by the Academy of Finland project ``Nestor'' (286211) and the EC H2020 RIA project ``SoBigData'' (654024). 
We would like to thank Chris Georgiev and the whole Imagga team for providing privileged API access for research purposes.


\bibliographystyle{SIGCHI-Reference-Format}
\balance
\small{
\bibliography{instagram_public_health}}

\end{document}